\documentclass[10pt,twocolumn,letterpaper]{article}

\usepackage{cvpr}
\usepackage{multirow}
\usepackage{times}
\usepackage{cite}
\usepackage{amsmath,amssymb,amsfonts}
\usepackage{algorithmic}
\usepackage{graphicx}
\usepackage{textcomp}
\usepackage{color}
\usepackage{comment}
\def\BibTeX{{\rm B\kern-.05em{\sc i\kern-.025em b}\kern-.08em
		T\kern-.1667em\lower.7ex\hbox{E}\kern-.125emX}}
\usepackage[ruled,vlined]{algorithm2e}
\usepackage{dsfont}
\usepackage{epsfig}
\usepackage{siunitx}
\usepackage{subfig}
\usepackage{mathtools}
\usepackage{booktabs}   
\usepackage{multirow}
\usepackage{bbm}
\usepackage{caption}
\usepackage[table,xcdraw]{xcolor}
\usepackage{pifont}
\usepackage{mwe}
\usepackage{cuted}
\usepackage{lipsum}
\usepackage[switch]{lineno}
\usepackage[T1]{fontenc}
\usepackage[utf8]{inputenc}
\usepackage{authblk}

\newcommand{\fref}[1]{Figure.~\ref{#1}}
\newcommand{\sref}[1]{Section.~\ref{#1}}
\newcommand{\tref}[1]{Table.~\ref{#1}}

\newcommand{\etandal}{\textit{et al}.}

\usepackage[pagebackref=true,breaklinks=true,colorlinks,bookmarks=false]{hyperref}

\usepackage[breaklinks=true,bookmarks=false]{hyperref}

\cvprfinalcopy 

\def\cvprPaperID{****} 

\begin{document}

\title{A Two-branch Neural Network for Non-homogeneous Dehazing via Ensemble Learning}

\makeatletter
\newcommand{\printfnsymbol}[1]{%
  \textsuperscript{\@fnsymbol{#1}}%
}

\author[1]{Yankun Yu \thanks{Authors contributed equally}}
\author[1]{Huan Liu \printfnsymbol{1}}
\author[1]{Minghan Fu}
\author[1]{Jun Chen}
\author[2]{Xiyao Wang}
\author[3]{Keyan Wang}
\affil[1]{Department of Electrical and Computer Engineering, McMaster University, Hamilton, Canada}
\affil[2]{Center for Research on Intelligent System and Engineering, Institute of Automation, University of Chinese Academy of Sciences, Beijing, China}
\affil[3]{State Key Laboratory of Integrated Service Networks, Xidian University, Xi'an, China}
\affil[ ]{\textit {\{yuy142,liuh127, fum16, chenjun\}@mcmaster.ca}, \textit{xiyaowang96@gmail.com},  \textit {kywang@mail.xidian.edu.cn}}

\renewcommand\Authands{ and }


\maketitle
\thispagestyle{empty}

\begin{abstract}
Recently, there has been rapid and significant progress on image dehazing. Many deep learning based methods have shown their superb performance in handling homogeneous dehazing problems. However, we observe that even if a carefully designed convolutional neural network (CNN) can perform well on large-scaled dehazing benchmarks, the network usually fails on the non-homogeneous dehazing datasets introduced by NTIRE challenges. The reasons are mainly in two folds. Firstly, due to its non-homogeneous nature, the non-uniformly distributed haze is harder to be removed than the homogeneous haze. Secondly, the research challenge only provides limited data  (there are only 25 training pairs in NH-Haze 2021 dataset). Thus, learning the mapping from the domain of hazy images to that of clear ones based on very limited data is extremely hard. 
To this end, we propose a simple but effective approach for non-homogeneous dehazing via ensemble learning. To be specific, we introduce a two-branch neural network to separately deal with the aforementioned problems and then map their distinct features by a learnable fusion tail. We show extensive experimental results to illustrate the effectiveness of our proposed method. The source code is available at \url{https://github.com/liuh127/Two-branch-dehazing}.
\end{abstract}

\vspace{-3mm}
\section{Introduction}
Single image dehazing as a low-level vision task has gained widespread attention in recent years. In the natural atmosphere, there are smoke, dust, haze, and other atmospheric phenomena that affect visibility. Pictures taken in these environments are often affected by blurring, color distortion, and low contrast problems.  
Using these kinds of pictures for classification, image segmentation, and other high-level vision tasks significantly reduces prediction accuracy.  Single image dehazing aims to restore a clean output image from a hazy input. Many dehazing methods \cite{dcp,li2017aod, cai2016dehazenet, zhang2018densely, Wu_2020_CVPR_Workshops, Qin_Wang_Bai_Xie_Jia_2020,  Liu_2020_CVPR_Workshops, zhou2020cggan, ren2018gated} have been proposed.

\begin{figure}[!t]
    \centering
    \includegraphics[width=0.5\textwidth]{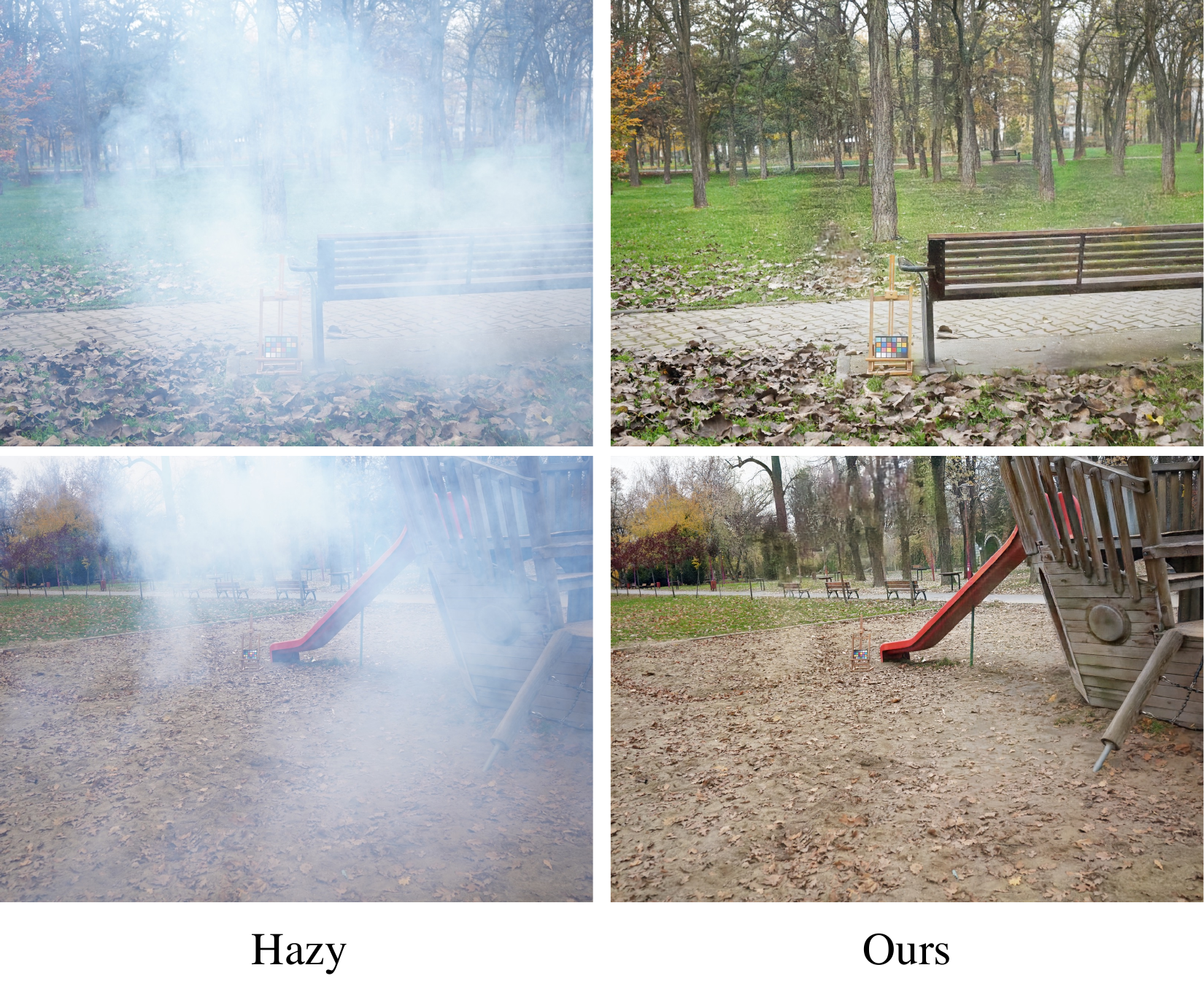}
    \caption{Qualitative results of our proposed method on NH-Haze 2021 validation set. Our method is able to produce haze-free image with high perceptual quality.}
    \label{fig:warp}
\end{figure}

The physical scattering model, \ie atmospheric scattering model (ASM) \cite{middleton1952vision}, provides the mapping formula between the hazy images and  clean counterparts.The ASM can be formally written as: 
\begin{equation}
I(x) = J(x)t(x) + A(1-t(x))
\end{equation}
$I$, $J$ respectively denote the hazy and haze-free image.  $A$ is the global atmosphere light.  $t(x)=$ $e^{-\beta d(x)}$ represents the transmission map, where $\beta$ and $d(x)$ are respectively the atmosphere scattering parameter and the scene depth. 

Many dehazing methods \cite{dcp, berman2016non, fattal2014dehazing,meng2013efficient} can achieve good performance based on ASM.  However, to generate a haze-free image with ASM,  dehazing models need to precisely estimate $t$ and $A$. As a result,  when dealing with images in complex environments, the inaccurate prediction of $t$ and $A$ usually leads to unsatisfactory dehazing results. 
Moreover, since $t$ depends on the distance of the scene, the atmospheric scattering model has a strong assumption, \ie the thickness of haze is strongly correlated to the depth of the background scene. This property forbids the ASM based methods to handle non-homogeneous hazy images. Somewhat surprisingly, recent years have witnessed the tremendous success of deep learning \cite{lecun2015deep} approach in addressing single image dehazing problem \cite{li2017aod,zhang2018densely, cai2016dehazenet, zhou2020cggan, Qin_Wang_Bai_Xie_Jia_2020,ren2018gated}. By using deep learning, the weakness of ASM can be avoided to a certain extent and the problem of exploring appropriate hand-crafted features is reduced to that of building a suitable convolutional neural network (CNN). With the availability of powerful CNNs, one can readily train them on large-scale datasets to learn a correct mapping from input hazy images to clear outputs.
However, it is costly (and in some cases impossible) to acquire vast quantities of hazy images with their corresponding clean ground truths in the real world. 

Recently, NTIRE organized several dehazing challenges and introduced several small-scale real-world datasets.
The challenge datasets have two intrinsic difficulties. 1) Limited training data. 
Using limited data, CNN based methods usually cannot gather enough statistical information of haze pattern, which results in a bad performance on dehazing. In other words, 
training on limited data is more likely to suffer from the over-fitting \cite{overfitting} problem. This problem is considered undesirable since it severely jeopardizes the generalization of models.
2) Complicated haze pattern. Most of the previous methods simply assume that haze is homogeneous. Since the non-uniformly distributed haze is more challenging to be removed than the homogeneous haze ( haze pattern cannot be simply formulated using ASM ), existing methods usually fail to produce satisfactory results when they work on non-homogeneous dehazing.


To cope with the complex distributions of non-homogeneous images and over-fitting problem in small-scale datasets, 
we here introduce our proposed method. To be specific, we design a simple two-branch neural network to deal with the aforementioned two issues separately.
The first branch, namely transfer learning sub-net, is built upon a ImageNet \cite{5206848} pre-trained Res2Net \cite{gao2019res2net}. 
ImageNet pre-training helps significantly alleviate over-fitting problems, especially before the large-scaled datasets are available to researchers \cite{he2019rethinking}. Besides, a pre-trained network is able to provide robust features for transfer learning \cite{Kornblith_2019_CVPR}. Therefore, the ImageNet pre-trained network is of great significance in solving the limited training data problem. However, we find that only using the ImageNet pre-trained model is not enough to tackle these issues. The pre-training on classification task usually fails to perfectly fit the target task, while a better solution is to find for data specific representations \cite{he2019rethinking}. To this end, we propose to add the other branch for fitting on current data, \ie current data fitting sub-net. This branch is trained from scratch and optimized only using the current training data. In favor of the strong mapping capability of residual channel attention network (RCAN) \cite{zhang2018image},  we build the current data fitting sub-net using RCAN. Unlike the original network setting \cite{zhang2018image} that down-samples the input images at the front of the entire network,  our second branch always maintains the original resolution of the inputs and avoids using any down-sampling operations. This adjustment avoids losing fine-detailed features. 
Finally, in order to aggregate the two varied outputs from our two branches, we design a fusion tail for learning a suitable ensemble strategy.

In summary, the main contributions of our work are as follows:
1) We demonstrate the effectiveness of using ImageNet pre-training in the non-homogeneous dehazing challenge\cite{ancuti2021ntire}. 
2) Towards learning data-specific representation, we propose to build current data fitting sub-net as a complement to the transfer learning sub-net. It can extract more distinctive features on current data distribution.
3) we adopt the idea from ensemble learning to design a learnable fusion tail. The fusion tail is simple and effective in fusing the outputs from two branches.
4) We show extensive experimental results to show the effectiveness of our two-branch network on both small-scale and large-scale datasets.

\section{Related Works}

\textbf{Single Image Dehazing}.
Single image dehazing methods can be roughly divided into two classes: prior-based methods and learning-based methods. Prior-based methods estimate the transmission map and the global atmosphere light in ASM \cite{middleton1952vision}. Many prior-based methods \cite{dcp,berman2016non, zhu2015fast} showed a good performance in single image dehazing. 
However, due to the prior is prone to violate in practice, the prior-based methods are not always robust when encountering complicated scenarios.   
With the success of deep convolutional neural networks, deep learning based methods received extensive attention in recent years. DehazeNet \cite{cai2016dehazenet} is the first deep learning based dehazing model. It adopts CNN to estimates the transmission map and then generate dehazed images using the physical scattering formulation. Unlike DehazeNet, AOD-Net \cite{li2017aod} is built to estimate both transmission and atmospheric light in one shot. In addition, many recent methods can recover haze-free images without using the physical scattering model. GFN \cite{ren2018gated} is a gated fusion network, which restores the hazy images with several transformations on the input, such as white balancing and gamma correction. GCANet \cite{GCA} employs smoothed dilated convolution layers to eliminate the gridding artifacts. Qin \etandal \cite{Qin_Wang_Bai_Xie_Jia_2020} proposed FFA-Net with novel feature attention modules, including pixel attention and channel attention. This work achieved high performance on the RESIDE \cite{li2019benchmarking} dataset. Shao \etandal \cite{shao2020domain} proposed a novel domain adaptation framework for dehazing tasks. This method bridges the gap between the real-world and synthetic hazy images. However, most of the previous methods perform bad when using real-world datasets introduced by NTIRE challenges, owing to the small-scale training set and the complex distribution of the haze. To solve the problems in real-world datasets, some approaches are proposed. Liu \etandal \cite{Liu_2020_CVPR_Workshops} introduced a Trident Dehazing Network(TDN) in the NTIRE2020 non-homogeneous dehazing challenge \cite{NTIRE_Dehazing_2020}. 
TDN consists of three branches using ImageNet pre-training, deformable convolution and many off-the-shelf techniques. Despite the remarkable achievements of TDN, the complicated network structure impedes us from carefully analyzing the significance of each technique.
Moreover, HardGAN \cite{deng2020hardgan} shows a good performance in both RESIDE \cite{li2019benchmarking} and NH-Haze 2020 \cite{Ancuti_2020_CVPR_Workshops, NH-Haze_2020} datasets. This method proposed a novel haze-aware feature distillation (HARD) module.  

\textbf{Ensemble Learning}.
Single output models are prone to suffer from the statistical problem, the computational problem, and the representation problem. 
Ensemble learning can partially solve these three problems \cite{dietterich2002ensemble}. It is based on the understanding that every model has limitations. Thus, ensemble learning aims to manage the strengths and weaknesses of single models. This management can make the best possible decision \cite{brown2010ensemble}. The ensemble architecture in our method can be classified as Mixtures of Experts \cite{jacobsadaptive}. 
The principle underlying the architecture is that both sub-nets in our model can focus on particular parts of the input space \cite{brown2010ensemble}. To generate a haze-free output, our fusion layer acts as a gating network that is responsible for learning the proper combination of the outputs from two branches.
Moreover, in our ablation studies, we demonstrate that this ensemble architecture fits well in our method.

\textbf{Transfer Learning}. Transfer learning is an effective way to solve problems with limited data. It aims to enable the system based on knowledge and skills learned in previous tasks to run on novel tasks \cite{pan2009survey}.  
The way we use transfer learning is based on the assumption that the network backbone is generalized and can extract versatile features \cite{tan2018survey}.
In our experiments, the network with pre-trained parameters surpasses the randomly initialized network by a large margin in terms of quantitative evaluation.

\begin{figure*}[!t]
\centering
\includegraphics[width=1.0\textwidth]{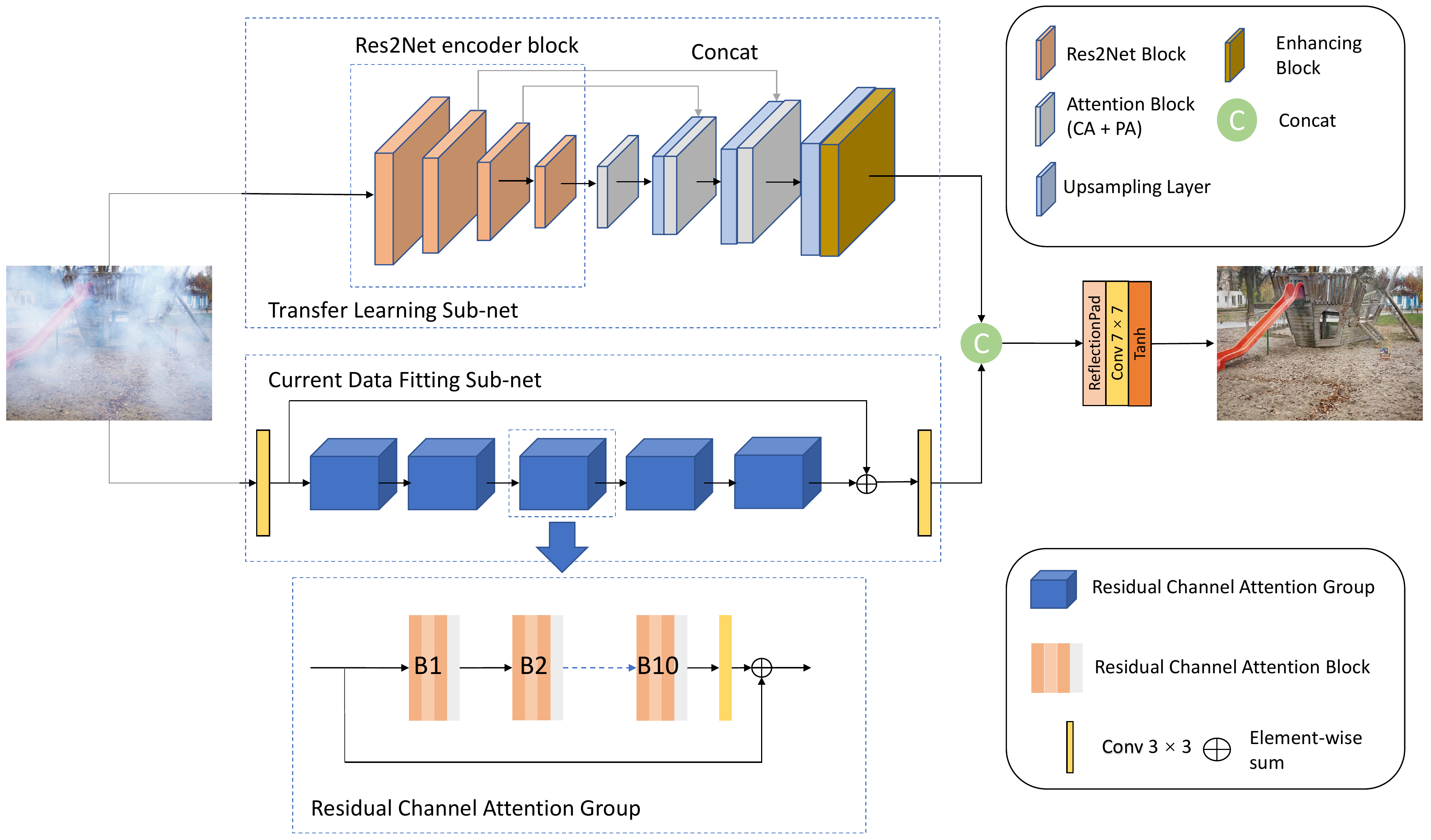}
\caption{Illustration of our model architecture.  }
\label{Fig.main}
\end{figure*}
\section{Proposed Method}
This section introduces the details of our two-branch neural network for non-homogeneous dehazing. Firstly,  we specify the details of the transfer learning sub-net and current data fitting sub-net. Then, we provide
the loss functions that are employed in the training stage.

\subsection{Network Structure}

As shown in \fref{Fig.main}, our method consists of two sub-nets, \ie the transfer learning sub-net and the current data fitting sub-net. Each sub-net is used for a specific purpose: transfer learning sub-net extracts robust global representations from input images with pre-trained weights, current data fitting sub-net aims to work on the current data and perform well on the specific training image domain. The fusion layer takes the concatenated feature maps of these two sub-nets and outputs haze-free images.

\textbf{Transfer Learning Sub-net.}
Transfer learning sub-net is an encoder-decoder network. Inspired by \cite{Wu_2020_CVPR_Workshops}, we use Res2Net as the encoder due to its excellent performance on the classification tasks. To be specific, we only adopt the front part of Res2Net with 16 times down-sampling and discard using the fully connected layer. Moreover, we adopt feature attention \cite{Qin_Wang_Bai_Xie_Jia_2020} as the attention module, and PixelShuffle \cite{shi2016real} as the up-sampling module in this sub-net. Before the features enter fusion tail, they will pass an enhancing module proposed by \cite{Qu_2019_CVPR}. At the training stage, the encoder module loads the ImageNet pre-trained parameters. Models with these pre-trained parameters can better extract robust features than those with randomly initialized parameters. However, as is pointed in \cite{he2019rethinking}, despite the effectiveness of using ImageNet pre-training in small-scale datasets, a better solution is to train the neural network on a large-scale dataset directly. Because accessing an identically distributed large-scale dataset is impossible during the NTIRE challenge, we alternatively consider taking benefits from specific data representation in small-scale challenge data to the utmost extent. Therefore, we build the current data fitting sub-net to achieve this object.
In addition, due to the nature of Res2Net, we can not apply a full-resolution skip connection from encoder to decoder. It may cause the network to lose some of the information that is important for restoring the image details. Consequently, we are further motivated to construct the second branch.

\textbf{Current Data Fitting Sub-net.}
Our current data fitting branch is based on residual channel attention block \cite{zhang2018image}. The block contains convolutional layers and channel attention modules. Owing to residual designing and long skip connections, the network is less likely to suffer from gradient vanishing problems \cite{pascanu2013difficulty}. Besides, channel attention highlights salient features to enhance the model's fitting ability on current data. Moreover, in order to preserve fine-detailed features, this sub-net avoids employing down-sampling and up-sampling operations. The fine-detailed features can be regarded as complements to that from transfer learning sub-net.  Since current data fitting sub-network is trained from scratch and built with full-resolution purpose, it would fit on the current data and extract more  specified features.  Nevertheless, the powerful fitting ability also makes it easier to encounter over-fitting problems ( see in \sref{sec:ablation} ). Thus, the role of transfer learning branch cannot be neglected.

\textbf{Fusion Tail.}
Our fusion tail as the ensemble method produces the final outputs. Specifically, the fusion tail takes the concatenation of features from two branches and then learns to map the features to clear images. The fusion tail is built with a convolutional layer followed by a hyperbolic tangent activation function. The reason we only adopt a single convolution operation is that the two branches have already provided sufficient and distinct features for restoring clear images.  In contrast, we find that a heavy fusion tail can potentially jeopardize the overall generalization ability of our method and results in performance degradation. To further illustrate the effectiveness of our fusion tail, we empirically study the effect of adopting fusion tails with different sizes in \tref{tab:tail}. It can be observed that with the increasing of fusion tail depth, the performance of our method degrades accordingly. 
\begin{table}[!t]
	\centering
	
    \setlength{\tabcolsep}{3.9mm}{
		\begin{tabular}{|c|c|c|}
			\hline
			\textbf{Fusion Tails} & \textbf{PSNR}&\textbf{SSIM}\\
			\hline
			three stacked residual blocks  &20.61 &0.8159\\
			\hline
			three convolution layers &20.72 &0.8148\\
			\hline
			Ours &\textbf{21.01}&\textbf{0.8195}\\
			\hline
	\end{tabular}}
	\caption{The performance of our proposed method using fusion tails with different sizes. The evaluation is conducted on NTIRE2021 validation set. Scores are provided by NTIRE2021 online server.}
	\label{tab:tail}
\end{table}

\subsection{Loss Functions}
Our loss function consists of four different components. Each one is used for a specific purpose.

\textbf{Smooth L1 Loss}.
We apply the smooth L1 loss to ensure the predicted images are close to clean images. It is a robust L1 loss \cite{Girshick_2015_ICCV} that is proved to be better than L2 loss in many image restoration tasks \cite{zhao2016loss}. 
\begin{equation}
    L_{l1} = \frac{1}{N}\sum_i^Nsmooth_{L1}(y_i - f_\theta(x_i))
\end{equation}
\begin{equation}
    smooth_{L1}(z) = \begin{cases}
0.5z^2 & \text{if }  |z| < 1\\
|z| - 0.5& \text{otherwise}
\end{cases}
\end{equation}
$y_i$ and $x_i$ denotes ground truth and hazy
image at pixel $i$. $f_\theta(\cdot)$ denotes our network
parametered by $\theta$. $N$ is the total number of pixels.

\textbf{MS-SSIM Loss}. 
In order to let the network learn to produce visually pleasing results, we adopt Multi-scale Structure similarity (MS-SSIM) as our second loss function. 
Let $O$ and $G$ denote two windows of common size centered at pixel $i$ in the dehazed image and the clear image, respectively. We apply a Gaussian filter to $O$ and $G$, and compute the resulting means $\mu_O$, $\mu_G$, standard deviations $\sigma_O$, $\sigma_G$, and covariance $\sigma_{OG}$. The SSIM for pixel $i$ is defined as

\begin{equation}
\begin{aligned}
    \text{SSIM}(i)&=\frac{2\mu_O\mu_G + C_1}{\mu_O^2 + \mu_G^2 +C_1} \cdot \frac{2\sigma_{OG} + C_2}{\sigma_O^2 + \sigma_G^2 +C_2} 
    =l(i) \cdot cs(i)
    \label{con:1}
\end{aligned}
\end{equation}
where $C_1$, $C_2$ are two variables to stabilize the division with weak denominator. $l(\cdot)$ denotes the luminance , and $cs(\cdot)$ refer to contrast and structure measures.

The MS-SSIM loss is shown in Eq. \ref{ms-ssim} where $\alpha$ and $\beta_j$ are default parameters. $M$ denotes the total number of scales.
\begin{equation}
    L_{\text{MS-SSIM}}(i) = 1- l_M^\alpha(i)\cdot\prod_{j = 1}^M[cs_j(i)]^{\beta_j}
    \label{ms-ssim}
\end{equation}

\textbf{Perceptual Loss}.
Unlike the MS-SSIM loss that focus mainly on the structural similarity, we adopt perceptual loss \cite{zhu2017unpaired} to provide additional supervision in high-level feature space. It has been acknowledged that training with perceptual loss allows the model to better reconstruct fine details. The loss network $\phi$ is VGG-16 \cite{simonyan2014very} that is pre-trained on ImageNet. The loss function is described as 
\begin{equation}
    L_{perc} = \frac{1}{N}\sum_{j}\frac{1}{C_jH_jW_j}||\phi_j(f_\theta(x)) - \phi_j(y)||_2^2
\end{equation}
where $x$ and $y$ are hazy inputs and ground truth images, respectively. $f_\theta(x)$ is the dehazed images. $\phi_j(\cdot)$ denotes the feature map with size $C_j\times H_j\times W_j$. The feature reconstruction loss is the $L_2$ loss. $N$ is the number of features that used in perceptual loss function.

\textbf{Adversarial Loss}.
Adversarial loss is proved to be effective in helping restore photo-realistic images \cite{ledig2017photo}. Especially for the small-scaled dataset, the pixel-wised loss function usually fails to provide sufficient supervision signals to train a network for recovering photo-realistic details.  Therefore, we finally implement the adversarial loss with the discriminator in \cite{zhu2017unpaired}. The loss function is described as
\begin{equation}
    L_{adv} = \sum_{n = 1}^N - logD(f_\theta(x))
\end{equation}
$D(\cdot)$ denotes discriminator. 
The probability that the dehazed image $f_\theta(x)$ is a ground truth image is shown as $D(f_\theta(x))$.

The total loss function is defined as:
\begin{equation}
   L = \gamma_1 L_{l1} + \gamma_2 L_{\textit{MS-SSIM}} + \gamma_3 L_{perc} + \gamma_4 L_{adv}
\end{equation}
where $\gamma_1$, $\gamma_2$, $\gamma_3$, and $\gamma_4$ are the hyperparameters to balance between different losses.

\section{Experiments}

In this section, we start with the description of datasets, training details, and evaluation metrics. Then we conduct ablation studies to clarify the effects of different modules in our method. Finally, we compare our method with other state-of-the-art dehazing algorithms quantitatively and qualitatively.
\subsection{Datasets} \label{sec:dataset}
We choose both real-world datasets and synthetic datasets to evaluate our network. For real-world datasets, we adopt the O-Haze \cite{8575270} from NTIRE2018 Dehazing Challenge \cite{8575287}, Dense-Haze \cite{Dense-Haze_2019,NTIRE_Dehazing_2019} in NTIRE2019 Dehazing Challenge \cite{dense-haze_report}, NH-Haze \cite{Ancuti_2020_CVPR_Workshops, NH-Haze_2020} used in NTIRE2020 Dehazing Challenge \cite{ancuti2020ntire}, and NH-Haze 2021 in NTIRE2021 Dehazing challenge\cite{ancuti2021ntire}. For synthetic datasets, we choose the Indoor Training set of RESIDE \cite{li2019benchmarking}.

O-Haze contains 35 pairs of outdoor hazy images and ground truth images for training. Dense-Haze,  NH-Haze 2020 and NH-Haze 2021 respectively contain 45 dense hazy images, 45 non-homogeneous hazy images, 25 non-homogeneous hazy images and their paired ground truths for training. Each of these datasets has 5 image pairs for validation and 5 pairs for testing.  In our experiments, we conduct our evaluation based on the official train, val and test split for O-Haze, Dense-Haze and NH-Haze 2020. For NH-Haze 2021, since the ground truth images for the validation set and testing set have not yet been released, we choose the first 20 official training pairs as our training data, and the rest 5 image pairs are used for evaluation. It should be noted that we conduct experiments on these datasets separately and do not use extra data to boost performance.

RESIDE is a benchmark for single image dehazing, which contains large-scale training and testing images in indoor and outdoor scenarios. The  Indoor Training Set (ITS) and the indoor Synthetic Objective Testing Set (SOTS) are used in our experiments.

\begin{figure*}[!t]
\centering
\includegraphics[width=1.0\textwidth]{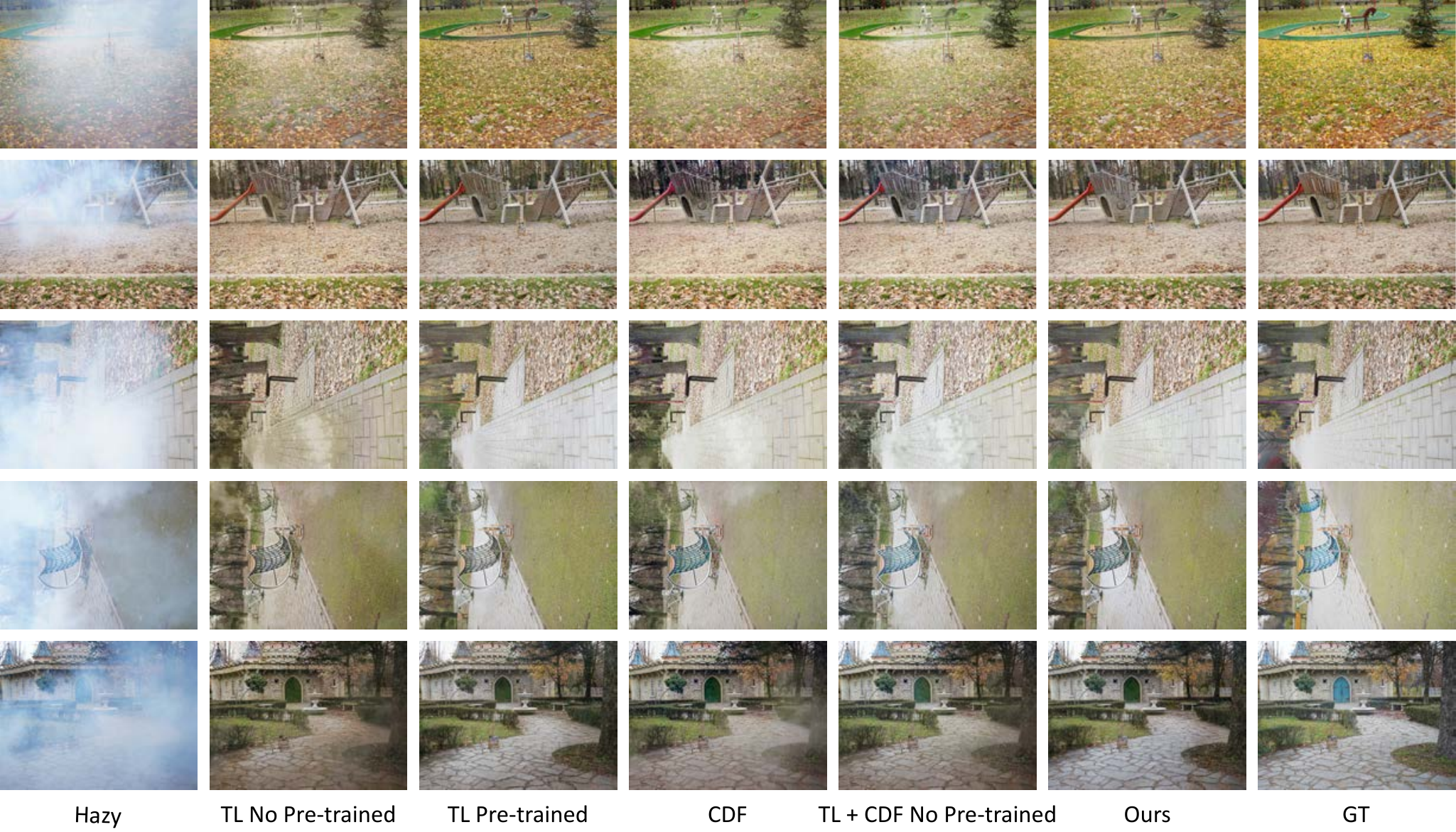}
\caption{The ablation study visual results of NTIRE2021 NH-Haze. }
\label{Fig.ablation}
\end{figure*}

\subsection{Training Details}
 We augment the training set with 90, 180, 270 degrees of random rotation, horizontal flip, and vertical flip. The input images are randomly cropped to a size of 256 $\times$ 256. We adopt Adam optimizer with default $\beta_1=0.9$ and $\beta_2=0.999$. The initial learning rate is 0.0001. The hyper-parameters of loss functions, $\gamma_1$, $\gamma_2$, $\gamma_3$, $\gamma_4$, are 1.0, 0.5, 0.01, 0.0005, respectively. Our  method  is  implemented  using  the  Pytorch  library \cite{paszke2017automatic}. All experiments are conducted on Nvidia V100 GPUs.
For quantitative evaluation, We adopt the Peak Signal to Noise Ratio (PSNR) metric and the Structural Similarity Index (SSIM) metric. 

\subsection{Ablation Analysis} \label{sec:ablation}

To intentionally analyze and evaluate the effectiveness of each architecture component, we conduct ablation studies by considering the combination of three factors: ImageNet pre-trained weights,  transfer learning sub-net, and current data fitting sub-net. The ablation experiments are shown as following:
1) TL without pre-trained weights: only use the transfer learning sub-net with randomly initialized parameters.
2) TL with pre-trained weights: only use transfer learning sub-net with ImageNet pre-trained weights.
3) CDF: only use current data fitting sub-net.
4) TL + CDF: use both current data fitting sub-net and transfer learning sub-net without ImageNet pre-trained weights.
5) Ours:  use both transfer learning sub-net with ImageNet pre-trained weights and current data fitting sub-net.

In detail, we use NH-Haze 2021 as the training set and testing set that respectively formed by first 20 images and the rest 5 images.  The quantitative results for ablation studies are shown in \tref{table.ablation}. From the table, we can observe that the ImageNet pre-trained weights can significantly improve the PSNR and SSIM of the transfer learning net. Although CDF is trained from scratch, it still out performs the TL without pre-training and approach to TL with pre-training in terms of PSNR.

On the other hand, since the current data fitting sub-net is built with full-resolution purpose, it would fit the current data and perform well on the specific training image domain. Thus, for the SSIM metric, the result of current data fitting sub-net is 0.062 higher than that of transfer learning sub-net with pre-trained weights.

All the two-branch models (TL+CDF and ours) perform better than those only with one branch. This indicates the effectiveness of our two-branch design.
It is no surprise that our method with all the components performs best. The PSNR and SSIM of our full model achieve 21.66 db and 0.843.  The scores indicate that every factor we consider plays an essential role in the network performance, especially the ImageNet pre-training.

\begin{table}[th]
\begin{center}
\setlength{\tabcolsep}{4mm}{
\begin{tabular}{|c|c|c|c|}
\hline
\multirow{2}{*}{\textbf{Methods}} & \textbf{ImageNet} & \multirow{2}{*}{\textbf{PSNR}} & \multirow{2}{*}{\textbf{SSIM}}\\
& \textbf{Pre-training} & & \\
\hline
TL & - &  17.92 & 0.555 \\
\hline
TL & $\surd$ &20.78 & 0.753 \\
\hline
CDF & - & 19.83 & 0.815\\
\hline
TL + CDF & - & 20.07 & 0.831\\
\hline
Ours & $\surd$ & 21.66 & 0.843\\
\hline
\end{tabular}}
\end{center}
\caption{Ablation study results. “TL” denotes transfer learning sub-net. “CDF” denotes current data fitting sub-net. "$\surd$ " denotes that the network is loaded with pre-trained weights, while "-" means no weights loaded.}
\label{table.ablation}
\end{table}

\begin{figure*}[th]
    \centering
    \includegraphics[width=0.8\paperwidth]{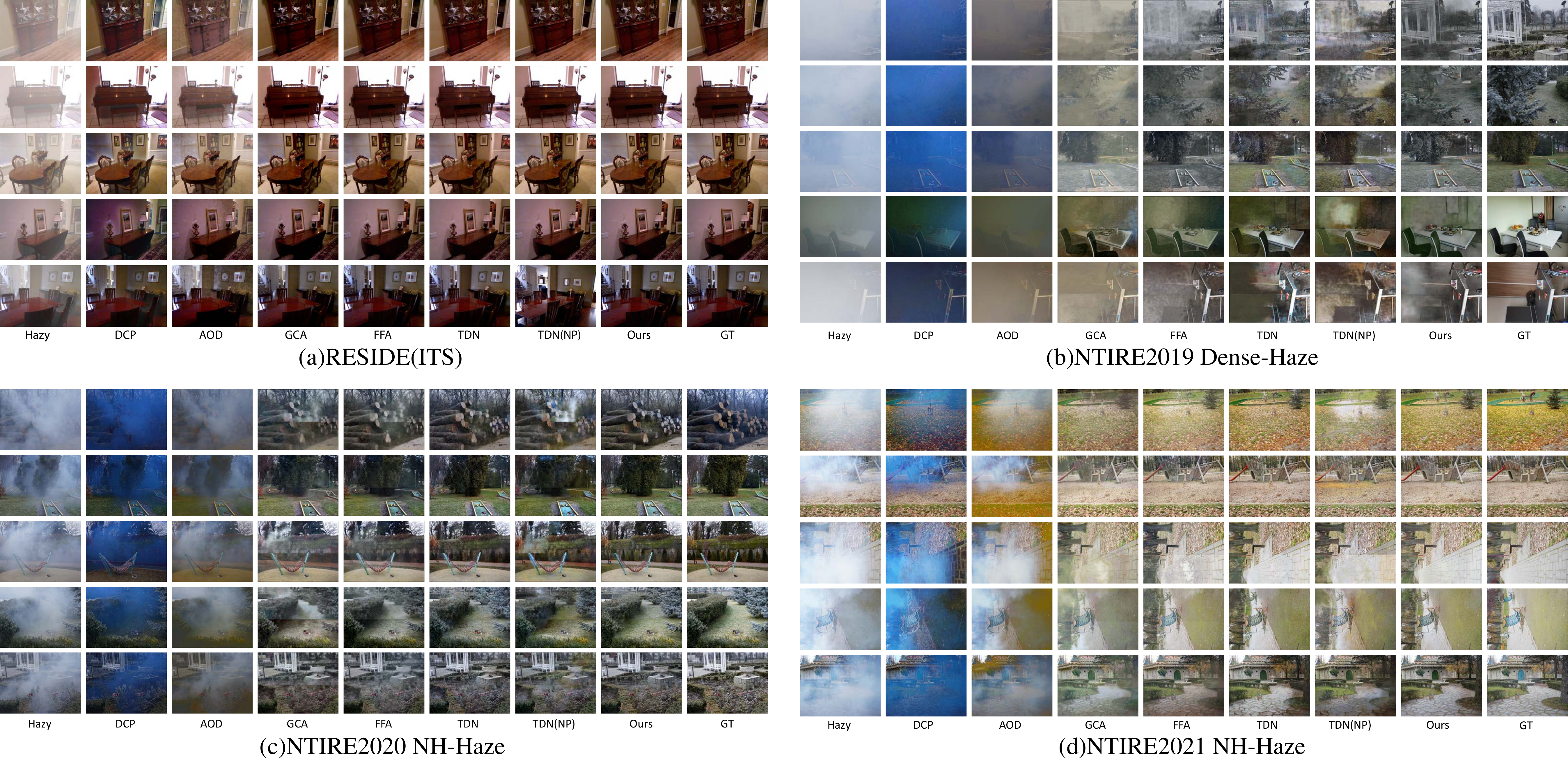}
    \caption{Qualitative evaluation on RESIDE(ITS), Dense-Haze, NH-Haze 2020 and NH-Haze 2021. }
    \label{fig:NTIRE}
\end{figure*}

\begin{table*}[t!]
\begin{center}
\setlength{\tabcolsep}{2.1mm}{}{
\begin{tabular}{*{12}{c|}}
\hline
 \multicolumn{2}{|c|}{\multirow{2}*{\textbf{Methods}}}& \multicolumn{2}{c|}{\textbf{RESIDE(ITS)}} & \multicolumn{2}{c|}{\textbf{NTIRE18(O-Haze)}} & \multicolumn{2}{|c|}{\textbf{NTIRE19}} & \multicolumn{2}{c|}{\textbf{NTIRE20}} & \multicolumn{2}{c|}{\textbf{NTIRE21}}\\
 \cline{3-12}
\multicolumn{2}{|c|}{} & \textbf{PSNR} & \textbf{SSIM} & \textbf{PSNR} & \textbf{SSIM} & \textbf{PSNR} & \textbf{SSIM} & \textbf{PSNR} & \textbf{SSIM} & \textbf{PSNR} & \multicolumn{1}{c|}{\textbf{SSIM}}\\
\hline
\multicolumn{2}{|c|}{DCP}& 16.62& 0.817& 12.92& 0.505& 10.85& 0.404& 12.29&0.411 & 11.30&0.605\\
\multicolumn{2}{|c|}{AOD}& 19.06& 0.850& 17.69& 0.616& 13.30&0.469 &13.44 &0.413 &13.22 &0.613\\
\multicolumn{2}{|c|}{GCANet} & 30.23& 0.975& 19.50& 0.660&12.42 &0.478 & 17.58 & 0.594 & 18.76 & 0.768\\
\multicolumn{2}{|c|}{FFA} &\underline{36.39} & \underline{0.988}& 22.12& \underline{0.768}& \underline{16.26}& \underline{0.545}& 18.51& 0.637 & \underline{20.40} & \underline{0.806}\\
\multicolumn{2}{|c|}{TDN}& 34.59& 0.975& \underline{23.53}& 0.754& 15.29&0.511 &\underline{20.51} &\underline{0.671} &20.31 &0.763\\
\multicolumn{2}{|c|}{TDN(No pre-trained)} & 30.52& 0.960& 21.67& 0.721& 15.10&0.495 & 17.29& 0.616& 17.73& 0.696\\
\multicolumn{2}{|c|}{Our} &\textbf{37.61} & \textbf{0.991}&\textbf{25.54}& \textbf{0.783}&\textbf{16.36} &\textbf{0.582} &\textbf{21.44} &\textbf{0.704} &\textbf{21.66} & \textbf{0.843}\\
\hline
\end{tabular}}
\end{center}
\vspace{-3mm}
\caption{Quantitative comparisons over SOTS-indoor, O-HAZE, Dense-Haze, NH-Haze 2020 and NH-Haze 2021 for different methods. The Best results are in \textbf{bold}, and the second best are with \underline{underline}.}
\label{table:1}
\end{table*}

\subsection{Comparisons with the State-of-the-art}
In this section, we show the comparisons between our method and the state-of-the-art. The comparison is conducted on the datasets introduced in \sref{sec:dataset}.
In our experiments, we select five state-of-the-art methods, including DCP \cite{dcp}, AOD-Net \cite{li2017aod}, GCANet \cite{GCA}, FFA \cite{Qin_Wang_Bai_Xie_Jia_2020}, and TDN \cite{Liu_2020_CVPR_Workshops}.
We also train a TDN without pre-training to verify the effect of ImageNet pre-training on this method. To be specific, we train all models on real-world datasets ( O-Haze, Dense-haze, NH-Haze 2020, 2021 ). On RESIDE dataset, we train AOD, TDN without ImageNet pre-trained model and ours from scratch. For the other methods, we use the results from their paper and released code.

\begin{figure*}[!th]
    \centering
    \includegraphics[width=1.0\textwidth]{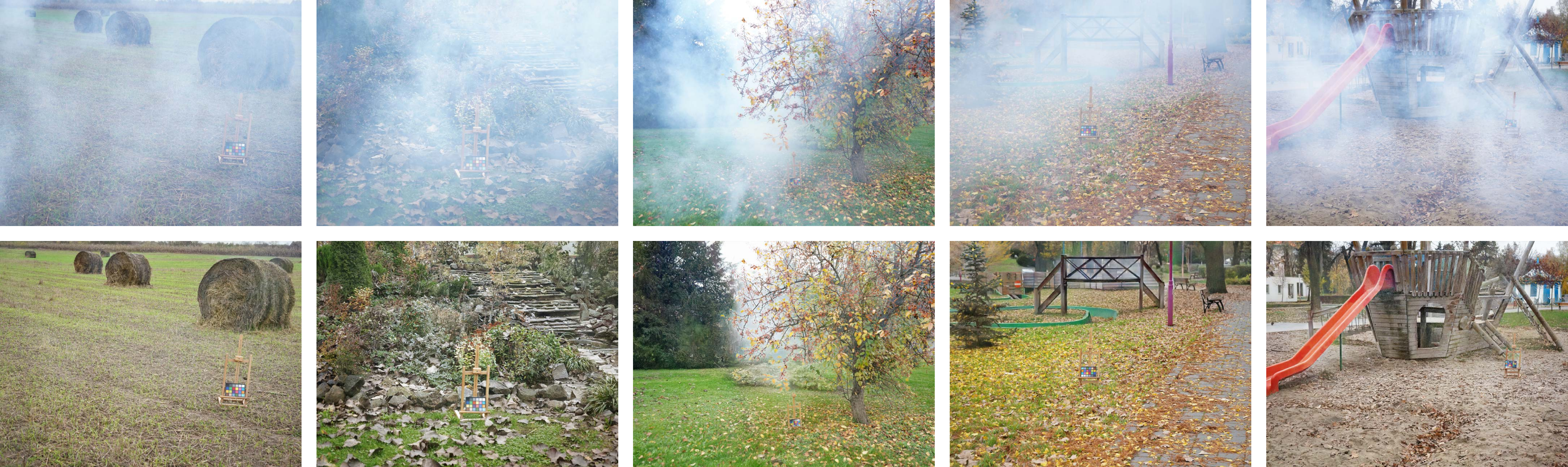}
    \caption{Test results of NTIRE2021 challenge. The fist row shows the original hazy images and second row illustrates dehazing results using our proposed method. }
    \label{fig:test_results}
\end{figure*}

\textbf{Results on Synthetic and Real-world Datasets}. The comparison results are shown in \tref{table:1}. Our method outperforms other methods by a significant margin 
in terms of PSNR and SSIM, especially on RESIDE(ITS), O-Haze, NH-Haze 2020 and NH-Haze 2021. For visual quality, we can observe from \fref{fig:NTIRE} that most of the methods can generate visually pleasing images on synthetic datasets except for DCP and AOD. In real-world datasets, DCP produces images with serious color distortion. AOD tends to generate dark images and can not remove the haze completely. The performance of GCANet on dense-haze is unsatisfactory. Its outputs still contain hazy areas and suffer from serious color distortion. Although FFA and TDN are able to generate better results than above methods, we can still find obvious visual problems, such as low brightness and blurry borders. For the comparison between TDN and TDN without pre-trained weights, we can conclude that the ImageNet pre-training significantly improves the performance when using small-scale NH-Haze 2020 and NH-Haze 2021 datasets. This further proves that the importance of using ImageNet pre-training as initialization when working on small-scale datasets. Additionally, in the last row of  \fref{fig:NTIRE} (d), all methods restore the blue gate to a green gate. However, the last image of \fref{fig:test_results} shows that our algorithm can successfully restore the blue gate.  One possible reason is that the images shown in \fref{fig:NTIRE} (d) are generated by the networks training on only 20 images. Limited training data can result in insufficient learning of color details.
It is also worth mentioning that images produced by our method are more visually pleasing. Our results are closest to ground truth images in terms of color and object details.

\begin{table}[th]
\begin{center}
\setlength{\tabcolsep}{4mm}{
\begin{tabular}{|c|c|c|c|}
\hline
\textbf{Methods} & \textbf{Parameters} & \textbf{Runtime}\\
\hline
AOD &  1.7K & 0.003s \\
\hline
GCA & 0.7M  & 0.255s \\
\hline
FFA & 4M & 0.561s\\
\hline
TDN & 46M  & 0.112s\\
\hline
TL & 49M & 0.038s\\
\hline
CDF & 1M & 0.033s\\
\hline
Ours(TL + CDF) & 50M  & 0.089s\\
\hline
\end{tabular}}
\end{center}
\caption{Parameters and runtime of each method.}
\label{table.complexity}
\end{table}

\textbf{Complexity Analysis}.
We run this experiment on a same NVIDIA V100 GPU. \tref{table.complexity} shows the total number of parameters and runtime per image (size 1600 $\times$ 1200) of above methods. Although the transfer learning(TL) sub-net contains a large number of parameters, its processing speed is quite fast. The reason is that the down-sampling operations shrink the size of features in the middle layer, which speeds up the computation. 
In conclusion, except for AOD-Net, the runtime of our method is faster than that of other state-of-the-art methods even though we have the most number of parameters.

\textbf{Results on NTIRE2021 Testing Set}.
During the NTIRE2021 dehazing challenge\cite{ancuti2021ntire}, we use NH-Haze 2020 as extra data. To eliminate the distribution shift between NTIRE2020 and NTIRE2021, we further employ gamma correction on NTIRE2020 with a hyperparameter of 0.65. 
Our results in NTIRE2021 test dataset are shown in \fref{fig:test_results}. The haze distribution in the original hazy images is non-homogeneous. Some areas have been turned purely white by the haze, while other parts can display the color of the grass field. In our dehazed images, we can see that in the first, second and last images, the haze is removed clearly. Although there is a small haze area remaining in the third and fourth images, the color and the outline of the leaves are clearly displayed. In the challenge report of NTIRE2021 \cite{ancuti2021ntire}, our dehazed results achieve 21.0183 and 0.8370 in PSNR and SSIM, respectively.

\section{Conclusion}
In this paper, we propose a two-branch neural network for non-homogeneous dehazing via ensemble learning and prove its strong power in various dehazing tasks. To generate images with fine detail and color fidelity, we stack the features from transfer learning sub-net and current data fitting sub-net and then map them to haze-free images by fusion tail. Our method has a significant advantage in small-scale datasets. It surpasses many state-of-the-art methods in both real-world and synthetic datasets. Besides, we demonstrate the effectiveness of the ImageNet pre-trained model.

{\small
\bibliographystyle{ieee}
\bibliography{egbib}
}

\end{document}